\documentclass{svproc}

\usepackage{tikz}
\usetikzlibrary{positioning}
\usepackage{url}

\AtBeginDocument{%
  \providecommand\BibTeX{{%
    \normalfont B\kern-0.5em{\scshape i\kern-0.25em b}\kern-0.8em\TeX}}}

\begin{document}
\mainmatter   

\title{A Fair and Ethical Healthcare Artificial Intelligence System for Monitoring Driver Behavior and Preventing Road Accidents}
\titlerunning{Healthcare Artificial Intelligence System for Monitoring Driver Behavior}

\author{Soraia Oueida\inst{1} \and Soaad Qahhār Hossain\inst{2} \and Yehia Kotb \inst{1} \and Syed Ishtiaque Ahmed \inst{2}}

\authorrunning{Oueida S. et al.}

\institute{College of Engineering and Technology, American University of the Middle East, Kuwait\\
\email{Soraia.oueida@aum.edu.kw and yehia.kotb@aum.edu.kw}
\and
University of Toronto, Canada\\
\email{soaad.hossain@mail.utoronto.ca and ishtiaque@cs.toronto.edu}
}

\maketitle    

\begin{abstract}
  This paper presents a new approach to prevent transportation accidents and monitor driver's behavior using a healthcare AI system that incorporates fairness and ethics. Dangerous medical cases and unusual behavior of the driver are detected. Fairness algorithm is approached in order to improve decision-making and address ethical issues such as privacy issues, and to consider challenges that appear in the wild within AI in healthcare and driving. A healthcare professional will be alerted about any unusual activity, and the driver's location when necessary, is provided in order to enable the healthcare professional to immediately help to the unstable driver. Therefore, using the healthcare AI system allows for accidents to be predicted and thus prevented and lives may be saved based on the built-in AI system inside the vehicle which interacts with the ER system.
 \end{abstract}


\section{Introduction}
Road traffic injuries and deaths have been and continue to be a major problem worldwide. Not only do they cost countries a noteworthy percentage of their gross domestic products (GDP), but they also, and more importantly those directly impacted from the road traffic injuries, incur a disability as a result of the injury. The Global Status Report on Road Safety 2018 by the World Health Organization (WHO) states that the number of road traffic deaths continues to rise steadily, reaching 1.35 million in 2016. Furthermore, the report states that road traffic injuries are estimated to be the eighth leading cause of death globally for all age groups, and the leading cause of death for children and young people 5–29 years of age \cite{ref1}. In terms of cost, in 2010 USD, it is estimated that fatal and nonfatal road traffic injuries will cost the world economy approximately \$ 1.8 trillion dollars from 2015–2030 \cite{ref2}. In addition, as Chen et al. said in their study on the global macroeconomic burden of road injuries, the health and economic burdens of road traffic injuries are distributed unequally across countries. In their study, they found that of the 70 million disability-adjusted life-years lost to road injuries worldwide in 2015, nearly 90\% occurred in low-income and middle income countries \cite{ref2}. This is especially problematic for individuals living in low-income countries as given that they are likely to lack good-quality hospital care \cite{ref2}. Individuals living in those countries who experience a vehicle accident may not be able to receive the care that they need, which can subsequently lead to them suffering from a disability or death. While previous traffic accident prevention studies have revealed that the availability and the robustness of data are usually the most problematic to the investigation of risk factors affecting the accident severity, there are empirical studies that show that most accidents were influenced by multiple behavioral factors \cite{ref3}. For instance, the study by Staubach found that while distractions is a driving error that causes road traffic accidents, two main groups can be identified: distractions caused by secondary activities and negative thoughts or emotions \cite{ref4}. However, even with knowing the behavioral factors that causes road traffic injuries, the problem is that in the wild, there is nothing in place that connects healthcare professionals to drivers experiencing those behavioral factors; the connection between the two is formed after the accident has taken place. Consequently, healthcare professionals are unable to intervene in a way that prevents the accident from occurring. Predicting unusual behavior of a driver such as speeding, drowsiness, high pulse rate, a possible stroke, etc. may prevent and decrease road accidents. A mobile application was developed by Kashevnik et al. to monitor and analyze drivers' behavior and predict dangerous states and thus alert the driver for safety purposes \cite{ref5}. 
\newline \newline
The healthcare artificial intelligence (AI) system presented in this work targets a medical interaction between the intelligent system integrated in the vehicle to detect unusual behavior and dangerous states with healthcare professionals inside an emergency room (ER). This interaction will help alerting the healthcare professional in case of a possible emergency and thus will lead to preventing an accident and providing safety by forwarding rescue to the corresponding driver and thus leading to a decrease in road accidents and mortality. The main contribution of this paper is the proposal of a fair and ethical AI-tool that could be used by healthcare and medical professionals to assist unstable drivers and prevent road accidents. What makes this healthcare AI system unique and different from other AI systems that use deep learning for driver behavior monitoring and accident prediction is that it takes into consideration fairness and ethics, both which are heavily valued and needed especially for healthcare and medicine, and is designed for use for healthcare and medical settings specifically. 
\newline \par
Vehicle's intelligence is based on the integration of the vehicle and the driver as a one unit. Here, the intelligent system needs to include but not limited to: deep learning, decision making, artificial intelligence, computing and many more network technologies. The purpose of this work is to propose an intelligent system that ensures the inter-connectivity with medical healthcare systems based on humans' abilities and vehicles' intelligence as well. Based on this inter-connectivity between medical devices and vehicles, healthcare professionals can monitor the driver in case of an alert signaled in real-time and are able to collect location data about the patient at risk and therefore provide immediate medical intervention which leads to a decrease in mortality rate. Along with this solution, a vehicle facing a problem may be urged to alert other drivers on the road by providing a certain alert that can be discussed later during the implementation phase, such as turning a red flasher, printing an alert message on the front window, etc. In this context all drivers will be alerted to avoid that specific vehicle leading to a decrease in the road accident rates.
\newline
\par\noindent
The paper is organized as follows: Section \ref{lit} provides an overview about the topics discussed in this paper through a detailed literature review. The healthcare AI system along with the potential solution are illustrated in Section \ref{system} including all the aspects of decision making and expands on the conception of what such AI system in healthcare should offer, machine learning and the interaction with medical professionals. Challenges and limitation are listed in Section \ref{limitations}. Finally, Section \ref{conc} concludes the paper and presents the potential future work.

\section{Literature Review}
\label{lit}
This section is dedicated for presenting the latest literature review on the main topics of this paper including but not limited to: Machine Learning, Artificial Intelligence, Driver Behavior Monitoring and their application to healthcare.

\subsection {Driver Behavior Monitoring}
Driving Behavior can be understood as the concept describing how the driver operates the vehicle in the context of the driving scene and surrounding environment \cite{ref15}. To understand the driver behaviors, majority of studies capture the driver status information, such as the head pose, eye gaze, hand motion, foot dynamics, and physiological signals \cite{ref6}. Physiological signals, such as the electroencephalogram (EEG) and electrooculography (EOG) are widely used for real-time driver status monitoring, which EEG signals are proved to be closely related to the driver behaviors, and can illustrate an earlier response to the mental states compared with the outer physical behaviors \cite{ref6}. Furthermore, as EEG can measure physiological responses to specific driving conditions, many researchers use EEG to detect risky driving behaviors \cite{ref7}. To elaborate, with using EEG, it allows for a high temporal resolution and direct record of neural activity \cite{ref8}. This high temporal resolution allows for a decomposition of the time domain EEG signal into spectral information through Fourier analysis, allowing for an examination of oscillatory activity in canonical frequency bands, which have been related to specific neuro-cognitive functions \cite{ref8}. These EEG signals generate a large amount of spatially oriented data over relatively short duration. Consequently, this leads to a big data problem \cite{ref11}. To overcome this, the WPCA-based feature reduction method by Dong et al. can be used, which doing so allows the dimension of EEG signals to be reduced for the purpose of using it for real-time processing needed for real-time diagnosis \cite{ref11}. 

\subsection {AI System Design, Machine Learning and Driver Behavior Prediction}
For the design of AI systems, it is important to note that the way people interact with statistical probability infused data-driven systems, such as AI systems, is not simple. Several studies have reported that individuals take their emotional and cultural values in concerns to connect to the machines while grouping with the probabilistic model-based tools and techniques and working side-by-side \cite{ref13}. In addition, many studies have shown that people experienced confusion after they failed to comprehend how the complex models and technologies worked, leading to mistrust in those systems \cite{ref13}. What resulted was literature within human-computer interaction (HCI), specifically that from Sultana et al., emphasizing that data-driven systems should be more engaged with other existing design paradigms that account for human values in a principled and comprehensive manner \cite{ref13}. The notion of AI and human values is not limited to HCI literature. Within AI literature, the goal of AI value alignment is emphasized - the goal that AI should properly align with human values \cite{ref14}. In aligning AI with human values, this will lead to an increase in trust from humans in AI systems, increasing the likelihood of individuals adopting AI. As such, in the wild, whether it is in a healthcare setting or legal setting, for the AI system to be adopted and optimal, the system must be designed in a way that enables it to account for human values. In the case of a healthcare AI system, the system should be designed in a way that accounts for patients, doctors, nurses and other healthcare professionals values. 
\newline \par
The concept of driver behavior has been studied within intelligent transportation systems (ITS) and human factors literature \cite{ref16}. Additionally, there has been multiple studies that investigated AI and its ability to predict dangerous states or unusual behavior for road accident and safety purposes. Previous studies investigated AI systems that utilize neural networks and other machine learning algorithms for detection of unsafe patterns and crash prediction \cite{ref12}. Learning algorithms in driver behavior detection and driving style studies are summarized in the study by Meiring and Myburgh \cite{ref18}. The study by McDonald et al. provides a comprehensive analysis of feature generation, machine learning, and input measures; specifically, the machine learning approaches that they studied were regression tree (RT), k-Nearest Neighor (kNN), support vector machine (SVM), neural network (NN), and convolutional neural network (CNN) \cite{ref20}. Of all the learning algorithms, artificial neural networks (ANN) are one of the most frequently used for unusual driving behavior \cite{ref15}. In addition, recurrent neural networks (RNN) have been used for driver behavior profiling \cite{ref19}. Streiffer et al. developed an approach that Internet of Things (IoT) and deep learning based approach to distracted drivers called DarNet - a multi-model data collection and analysis system designed to detect and classify distracted driving behavior, which their paper discusses about the system design and implementation \cite{ref17}. Gao et al. developed a spatial–temporal CNN that uses EEG signals to detect driver fatigue \cite{ref9}. Similarly, Xing et al. used CNN and transfer learning for an end-to-end driving related activities recognition system, which the system could be used for driver behavior monitoring. Their system involved the use of a CNN model, Gaussian mixed model (GMM), AlexNet model, and RGB images obtained from an RGB camera \cite{ref21}. These, along with many other studies, have shown that machine learning and deep learning algorithms can be used to analyze non-visual features and visual features with respect to driver behavior and driving activities \cite{ref16}, making them efficient in predicting unusual driver behavior and driving activities. However, despite there being multiple studies, many which show how machine learning is efficient in predicting dangerous states or unusual behavior for preventing road accidents and providing road safety, the application of machine learning models in driver behavior analysis is still limited and more effort is needed to obtain well-formed and generalizable results \cite{ref15}. Our paper contributes to this effort of furthering the effort for well-formed and generalizable results as it will utilize a machine learning approach that can be applied to driver behavior and activities, and be used by a system that supports healthcare professionals. 

\section{The Healthcare AI System}
\label{system}
\par\noindent The majority of accidents happen as a result of mistakes drivers do. If there is a way to detect those behaviors that lead to risks, then, accidents could be avoided and risks could be eliminated. The following subsections discuss the proposed model for detecting and reporting those situations.  
\subsection {The System and Machine Learning}
\par\noindent The problem of detecting driver bad behavior is not an easy task since it depends on the individual behavior that differs from a driver to another. This means that the system needs to learn driver behaviors in two phases. The first phase is the static phase which is a supervised learning and during this phase, the system learns the common mistakes and bad behavior that all drivers do. The second phase is a dynamic learning phase which learns customized behavior of the driver and this phase is unsupervised. The actions that are being selected after recognizing the bad behavior of the driver is based on reinforcement learning. The below figure describes the proposed system.

\begin{figure}[htp]
    \centering
    \includegraphics[width=9cm]{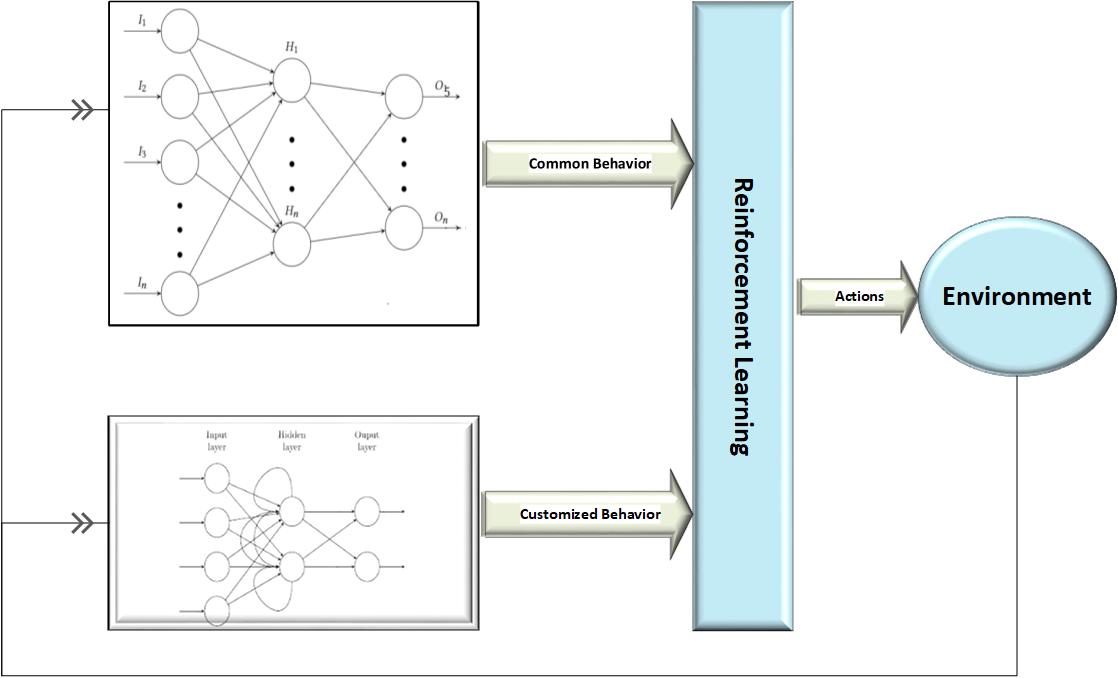}
    \caption{AI based behavior learning}
    \label{fig:behaviorLearning}
\end{figure}

\par\noindent As seen in Figure \ref{fig:behaviorLearning}, after the system learns, the observed situation is fed to the neural networks and the neural network notifies whether the situation is classified as dangerous or not. The recognized situation is fed to the reinforcement system to select an action to be done. \newline
\par\noindent ANN is to be used for the supervised learning to learn about common unusual driving behavior. This is basically because the pattern is already known and we can provide a training set that specifies the situation with the class it belongs to. The customized bad behavior however, depends on the individual and therefore it needs to be unsupervised. An RNN is used for clustering those behaviors. The system will cluster the behaviors and recognize different classes and based on the class the action is determined. Choosing an action depends on the recognized situation whether it is common or customized. This is done through reinforcement learning with a reward function that increases when the bad behavior effect is avoided. \newline

\par\noindent Driver bad behavior could result to accidents. Bad behavior could also be due to health issues such as drowsiness and diabetes. High blood sugar levels effect concentration and the ability to do better judgements. The system requires sensors such as wearable devices and gauges to feed the system with readings to predict any future possibility of a bad behavior. Also, this helps notifying emergency units about reading anomalies and reporting any urgent actions needed. It is worth noting that the system will work regardless what kind of gauges connected to it. In other words, it does not have to be diabetes. Any health issue that can be detected by a gauge can be provided as an input parameter to the system. Notification to emergency departments can be achieved through wireless network protocols. 


\subsection{Algorithmic Fairness, Ethics and Decision Making}
Fairness is increasingly recognized as one of the most critical components of AI and machine learning systems \cite{ref24}. As such, we will elaborate on how our AI system achieves fairness to an acceptable manner. One major issue that existed with machine learning approaches that relied on visual features is that they relied on computer vision, which computer vision suffers from its own set of issues. For instance, in 2018, it was found that Amazon’s facial recognition system, called Rekognition, failed to accurately recognise the faces of darker-skinned women close to 30 percent of the time \cite{ref23}. This is a major issue as Amazon's AI system did not align human values. Consequently, this led to trust issues between the users and that system, especially from Black individuals and people of color. Until the issue of recognition within computer vision has properly and fully been addressed, there is always the possibility that AI systems that use computer vision for driver behavior monitoring can encounter racial and other recognition-related issues. In using non-visual features such as heart rate, our healthcare AI system is able to avoid racial and other recognition issues in a way that allows for black, indigenous and people of color (BIPOC) to be addressed in the same manner as white individuals. With using health data such as heart rate, issues found within computer vision can be avoided. \newline \par

When an idea that involves using a person's location and personal information is proposed, the concerns around privacy and security follow, which is fair as privacy and security is something that is of high value to people and data breaches and other such issues is a growing concern with technologies. Given that driver behavior monitoring systems use a person's location and personal information, there is always the possibility that driver behavior monitoring systems could be used against the user through having the AI-tool used as a surveillance-tool. With our healthcare AI system, as there will be a feature that allows driver's locations to be revealed to the healthcare professional, we acknowledge that the concern of it being used as a surveillance tool is an issue that applies to our system, and that must be addressed. The way in which we plan to address this is through decision-making and the state of the unstable driver. Simply put: only when the driver is in a life-threatening situation will their location be provided to the healthcare professionals. For instance, if the intelligent system integrated in the vehicle detects a very high dangerous state and the medical professionals are not able to communicate with the driver, then the professionals can send an ambulance and paramedics to the driver. If the unstable driver gets into a road accident, then the driver's location will be provided to the healthcare professionals so that the professionals can send an ambulance and paramedics to the driver. The idea for this is similar to notion of airbags within cars in the sense that just as airbags are only activated when a major car crash occurs, the location of the driver is only activated when a major road accident occurs. For both cases presented, the intelligent system in the vehicle makes the decision of providing the driver's location, not the healthcare professional. Consequently, professionals such as those in law enforcement units cannot use the healthcare AI system for surveillance and other purposes. 

\subsection{Interaction Between Healthcare Professionals and Healthcare AI System}
Earlier in the paper, we said how our healthcare AI system targets a medical interaction between the intelligent system integrated in the vehicle to detect unusual behavior and dangerous states with healthcare professionals inside an ER. Figure 2 is a diagram showing the interaction between the unstable driver, healthcare AI application and the healthcare professionals. 

\begin{figure}[htp]
    \centering
    \includegraphics[width=9cm]{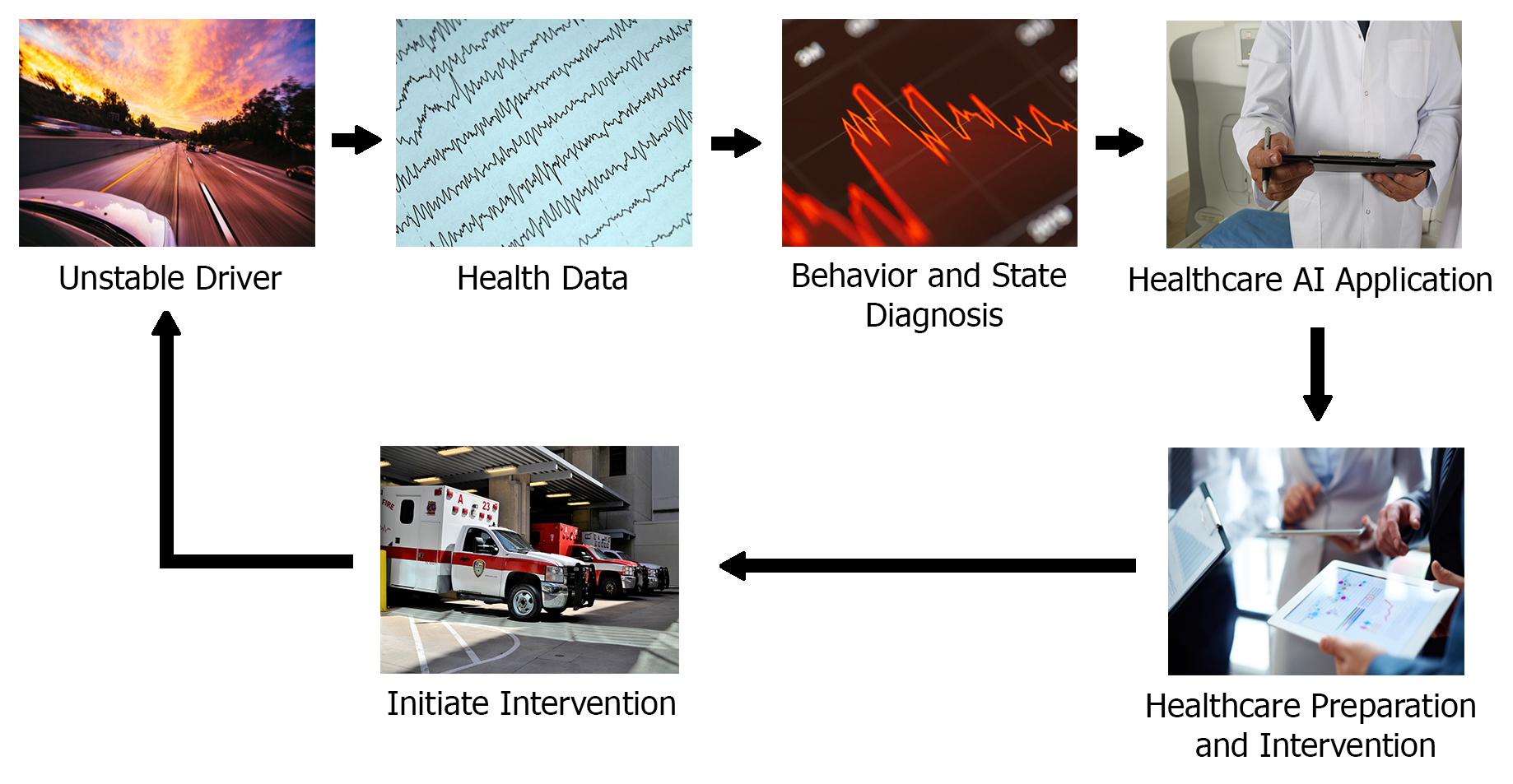}
    \caption{Healthcare AI system}
    \label{fig:healthai}
\end{figure}

The intelligent system integrated in the vehicle monitors health data, such as heart rate, of the driver. As such, while the exact approach for how the intelligent system will obtain the health data of the driver has not been fully established presently, an approach that involves sensors is being considered. After the health data has been collected and analyzed by the intelligent system, the system will then decide whether to notify the healthcare professionals in the ER through the healthcare application. Once the intelligent system has decided to notify the healthcare professionals, the system will send notifications and share all relevant details regarding the unstable driver to the healthcare AI application. A triage or registered nurse will interact with the healthcare AI application to view the notifications and review and assess the information regarding the driver. The reason why a triage or registered nurse will interact with the application first and not an emergency physician is because in the wild, emergency physicians are constantly on the move and working from one case to another. It would be difficult for an emergency physician to deal with the healthcare AI application while attending to their other responsibilities and commitments. Knowing this, the design of the healthcare AI application would be one that is nurse-friendly. That is, it will incorporate values and features that will help the triage or registered nurse feel comfortable and approach situations presented by the AI in an optimal manner. Upon assessing the information, the triage nurse will involve the appropriate healthcare or medical professional. If the case is one that involves the driver feeling highly distressed or suicidal for example, the triage or registered nurse will get a social worker or another mental health professional to get in touch with the driver through the healthcare AI application. If the case is one that has the driver being in a road accident, then the triage or registered nurse will contact the paramedics to retrieve the driver for medical assistance and the emergency physicians, hospital medical officers, nurse practitioners, etc. to prepare for the treatment of the driver. If the case is where the driver is drunk, then the triage or registered nurse will contact a taxi and a tow truck to drop the driver and his or her car back to his or her home, and the triage or registered nurse will get the intelligent system to communicate the plan to the driver. Alternatively, the triage or registered nurse can also speak to the driver to let the driver know of the plan. Who speaks to the driver will depend on the settings set by the driver. Some drivers may prefer and be more comfortable speaking to an AI system, while others may feel more comfortable speaking to a person. These preferences along with other preferences will be integrated within the healthcare AI system so that the system aligns with the values of the drivers, but also enables the healthcare professionals to assist the driver when needed. Similarly, the healthcare AI application will be designed in a way that meets the needs and demands of the emergency physicians, triage nurses, social workers and other healthcare professionals in the ER. However, the design will not necessarily be the same as in the wild, different healthcare professionals require different tools and information in order to operate. For instance, the interface that a social worker would use will be different than that of a triage nurse as the approach that a social worker would use to assist an unstable (e.g. distressed) driver would be different than that of a triage nurse. In addition, the social worker may need tools and data that the triage nurse may not need, such as a virtual notepad and elaborate details on the mental state of the driver. \newline \par

As much as we want to keep the healthcare AI system focused on the medical interaction and assisting the driver with their health-related needs and safety, we cannot due to how things are in the wild within driving settings. In the wild, health and safety are not the only priority of drivers; legal and financial concerns are of a priority for drivers as well. As such, we are considering measures that protect the driver from getting into legal trouble and financial issues, and measures to protect surrounding drivers and pedestrians. The idea of having intelligent system within the vehicle activate the back lights, in a non-red color, of the unstable driver's car from the initiation of the medical interaction to the conclusion of it is an idea that is being considered. The motivation for this is to let surrounding vehicles and pedestrians know to distance themselves from the vehicle containing the unstable driver, and to let law enforcement units such as the police know that a medical interaction is taking place, and that they should not intervene unless if its the case that the road accident has taken place while they were around. In that case, they can intervene and help ease the situation and control traffic on the roads. Additionally, the police can interact with the driver or the intelligent system to confirm whether paramedics are on their way. 

\section{Challenges and Limitations}
\label{limitations}
The major challenge in the context of this project is the availability and accessibility to rich medical data. Data must include patient's information such as age, sex, medical history, etc. The challenges must be related to the prediction phase and the accuracy of the machine learning in the decision making process.  During implementation, full cooperation of some healthcare systems is needed along with the ability of testing some vehicles and therefore the cooperation of some drivers along the journey. Another challenge, which applies more for driver behavior monitoring in general, the success of behavioral prediction is very low due to the challenges associated with analyzing non-visual features and visual features with respect to driver behavior and activities in the wild \cite{ref17}. When we say in the wild, we mean driver behavior and driving activities that take place in a real-life, day-today setting, and not driver behavior and driving activities that take place through virtual reality (VR) or some setting that has been modified for studying purposes.

\section{Conclusion}
\label{conc}
The advancements in the technology led professionals and researchers improve the available networks in order to include intelligent systems able to provide more insights on accessible data, promote services and efficiently improve many sectors. One of the main benefit in this concern is to improve the transportation field where road accidents can be avoided and lives can be saved. The intelligent system discussed in this paper must be implemented inside the vehicle in order to help preventing road accidents by alerting other vehicles through signaling an alert and must be integrated with a healthcare department such as an ER to allow healthcare professionals to interfere in case of an emergency medical case impacting the driver such as a high blood pressure, a possible stroke, a bad vision, drowsiness, etc. The intelligent vehicle can provide the healthcare system with the current location of the driver, or connect the driver to a healthcare professional such as a social worker or other mental health expert, in order to receive the appropriate medical care. The development of the solution proposed will be the next phase of this work, where data must be collected to start the learning process and feed the system with enough information to be able to precisely predict any miss-behavior of the driver and therefore perform the required actions using the intelligent system proposed. More factors will be investigated during the implementation phase of this project including, but not limited to, the right learning algorithm to be used for machine learning and decision making, the type of data needed to enrich the learning and the hardware to be accessible. \newline \par

As a future work, the communication among the intelligent vehicles along with the interconnection of these vehicles with healthcare systems as discussed in this paper, will be developed to ensure the ability to produce a well established artificial intelligence integration for monitoring driver's behavior and preventing road accidents for both low-income cities and smart cities. The implementation of IoV (Internet of Vehicles) allows vehicles connected to the same smart city infrastructure to communicate in order to produce an alert in case of any driver bad behavior. Therefore, some challenges are expected to occur when implementing IoV and IoMT (Internet of Medical Things) in the smart city where connectivity and networking of vehicles along with humans as one whole entity must be maintained. 

\bibliographystyle{plain}

 \bibliographystyle{plain}
\bibliography{ftc}

\end{document}